\documentclass[submission,copyright,creativecommons]{eptcs}
\providecommand{\event}{SOS 2007} 

\usepackage{iftex}
\usepackage[section]{placeins}
\usepackage{amsthm, amssymb}
\usepackage{graphicx}
\usepackage{tikz}
\usepackage{pgf}
\usepackage{tikz-3dplot}
\usetikzlibrary{positioning}
\usetikzlibrary{arrows}
\usetikzlibrary{shapes}
\theoremstyle{definition}
\newtheorem{definition}{Definition}[section]

\ifpdf
  \usepackage{underscore}         
  \usepackage[T1]{fontenc}        
\else
  \usepackage{breakurl}           
\fi

\title{Privacy-preserving Linear Computations in Spiking Neural P Systems}
\author{Mihail-Iulian Plesa
\institute{University of Bucharest\\ Bucharest, Romania}
\institute{Department of Computer Science}
\email{mihail-iulian.plesa@s.unibuc.ro}
\and
Marian Gheorghe 
\institute{University of Bradford\\ Bradford, UK}
\institute{School of Electrical Engineering and Computer Science}
\email{M.Gheorghe@bradford.ac.uk}
\and 
Florentin Ipate
\institute{University of Bucharest\\ Bucharest, Romania}
\institute{Department of Computer Science}
\email{florentin.ipate@unibuc.ro}
}
\def\titlerunning{On Privacy in Spiking Neural P Systems}
\def\authorrunning{M.I. Plesa, M. Gheorghe \& F. Ipate}
\begin{document}
\maketitle

\begin{abstract}
Spiking Neural P systems are a class of membrane computing models inspired directly by biological neurons. Besides the theoretical progress made in this new computational model, there are also numerous applications of P systems in fields like formal verification, artificial intelligence, or cryptography. Motivated by all the use cases of SN P systems, in this paper, we present a new privacy-preserving protocol that enables a client to compute a linear function using an SN P system hosted on a remote server. Our protocol allows the client to use the server to evaluate functions of the form $t_1k + t_2$ without revealing $t_1, t_2$ or $k$ and without the server knowing the result. We also present an SN P system to implement any linear function over natural numbers and some security considerations of our protocol in the honest-but-curious security model.  
\end{abstract}

\section{Introduction}
Membrane computing (or P systems) is a new model of computation inspired by how membranes work and interact in living cells \cite{puaun2000computing}. There are several variants of the model e.g. neural P systems, cell P systems, tissue P systems, etc.,  \cite{zhang2010survey,pan2010computational,ionescu2006spiking}. P systems have generated new perspectives on the P vs NP problem, being used to efficiently solve hard problems \cite{zandron2001solving,csuhaj2022membrane,fan2021implementation,zhang2008quantum}. There are also multiple applications of P systems in various fields like formal verification, artificial intelligence, or cryptography  \cite{zhang2017real}. 

In this work we used a special type of P systems called Spiking Neural P systems (SN P systems for short)  \cite{ionescu2006spiking}. SN P systems are inspired by biological neurons. There are also numerous variants of SN P systems: SN P systems with astrocytes, SN P systems with communication on request, SN P systems with polarization, SN P systems with colored spikes, etc.,  \cite{pan2012spiking,pan2017spiking,wu2017spiking,song2017spiking}. 

Although there are many theoretical aspects and simulations in the literature, to gain the maximum efficiency of these systems, they must be implemented on dedicated hardware  \cite{arteta2021vivo}. If these systems are implemented at a large scale, they will have to be accessed remotely in the cloud. This raises privacy concerns about data uploaded to the server that hosts the P system. This paper approaches the problem of confidentiality in SN P systems by describing a protocol that allows a client to perform a simple linear computation using an SN P system that is served remotely without revealing private information.

\subsection{Related work}

Besides the theoretical work, there are also many applications of P systems. In \cite{plesa2022key} the authors propose a new key agreement protocol based on SN P systems. In  \cite{ganbaatar2021implementation,guo2016family,guo2017implementation} the authors describe how to implement the RSA algorithm in the framework of membrane computing. One ingenious way of applying P systems is shown in  \cite{vasile2023breaking} which presents an algorithm to break the RSA encryption. There are also applications in artificial intelligence.  In \cite{chen2021survey} the authors present a survey of the learning aspects in SN P systems. Clustering algorithms have also been developed in the framework of membrane computing   \cite{peng2015automatic,peng2015unsupervised,peng2015novel}. Image processing is another common application of P systems  
\cite{diaz2019membrane,yahya2016image,diaz2013parallel}.

\subsection{Our contribution}

In this paper, we present a protocol that allows a client to perform a linear computation using an SN P system hosted on a server without revealing any private data. The SN P system computes functions of the form $t_1k + t_2$ over natural numbers. The client must retrieve from the server the result of the computation without the server knowing $t_1, t_2$ or $k$. Also, the server must not learn the value $t_1k + t_2$. To enable privacy-preserving computations on the server side, we use the ElGamal cryptosystem and its homomorphic properties \cite{elgamal1985public}. We also provide an SN P system that computes any linear function over natural numbers and some security considerations of our protocol. The paper is organized as follows: in Section 2 we present the background on the SN P system and homomorphic encryption. In Section 3, we show an SN P system that computes linear functions over the natural numbers. In Section 4, we introduce our protocol and some security considerations. Section 5 is left for the conclusions and further directions.  

\section{Preliminaries}
In this section, we briefly present the Spiking Neural P systems (SN P systems) and the cryptographic algorithm used in our protocol. We stress some useful properties of the encryption scheme. 
\subsection{Spiking Neural P systems}
A Spiking Neural P system (SN P system) of degree $m \ge 1$ is defined as the following construct:
\begin{definition}
$\Pi = \left(O, \sigma_1, \sigma_2,\dots,\sigma_m, syn, i_0\right)$ where:
\begin{itemize}
    \item $O=\{a\}$ is the alphabet. The symbol $a$ denotes a spike.
    \item $\sigma_i,\ 1 \le i \le m$ represents a neuron. Each neuron is characterized by the initial number of spikes denoted by $n_i \ge 0$ and the finite set of rules denoted by $R_i$: $\sigma_i=\left(n_i, R_i\right)$.
    \item Each rule can be of the following two forms:
    \begin{enumerate}
        \item $E/a^r \rightarrow a;t$ where $E$ is a regular expression over the alphabet $O$, $r \in \mathbf{N^*}$ represents the current number of spikes in the neuron and $t \ge 0$ is the refractory period. This type of rule is called a firing rule.
        \item $a^s \rightarrow \lambda$ where $s \ge 1$ is the current number of spikes in the neuron and $\lambda$ is a special symbol that denotes an empty set of spikes. This type of rule is called a forgetting rule.
    \end{enumerate}
    \item $syn \subseteq \{1,2,\dots,m\} \times \{1,2,\dots,m\}$  is the set of synapses between neurons. No neuron can have a synapse to it i.e. $\left(i,i\right) \notin syn$ $\forall\  1 \le i \le m$.
    \item $i_0$ represents the output neuron.
\end{itemize}
\end{definition}

A neuron can fire using the firing rule $E/a^r \rightarrow a;t$ only if it contains $n$ spikes such that $a^n \in L\left(E\right)$ and $n \ge r$ where $L\left(E\right)$ is a language defined in the following way:
\begin{itemize}
    \item $L\left(\lambda\right) = \{\lambda\}$
    \item $L\left(a\right)=\{a\}\ \forall a \in O$
    \item $L\left(\left(E_1\right)\cup \left(E_2\right)\right)=L\left(E_1\right) \cup L\left(E_2\right)$
    \item $L\left(\left(E_1\right)\left(E_2\right)\right)=L\left(E_1\right)L\left(E_2\right)$
    \item $L\left(\left(E_1\right)^+\right)=L\left(E_1\right)^+$
\end{itemize}
for all regular expressions over the alphabet $O$.

After firing, $r$ spikes are consumed. A firing rule that is applied when the neuron contains exactly $r$ spikes i.e. $L\left(E\right)=\{a^r\}$, is simply denoted as $a^r \rightarrow a;t$.

At the neuron level, all rules are applied sequentially, but the system as a whole evolves with maximum parallelism i.e. if a rule can be applied in a neuron then that rule will be applied. 

At a certain point, a neuron can be firing, spiking, or closed. If a neuron applies the firing rule $E/a^r \rightarrow a;t$ at moment $q$ then the neuron will send a spike to all the neurons to which it is connected by synapses at moment $q+t$. At times $q+1, q+2, \dots q+t-1$ the neuron will be in the refractory period i.e. the neuron will not receive or send any spikes. When neuron $\sigma_i$ is spiking, the spikes are replicated in such a way that each neuron $\sigma_j$ with $\left(i,j\right) \in syn$ receives one spike although the number of spikes consumed by $\sigma_i$ is exactly $r$.

When a forgetting rule $a^s \rightarrow \lambda$ is applied in a neuron, $s$ spikes are removed from that neuron. A neuron can apply a forgetting rule only if the number of spikes is exactly $s$.

There are several ways in which we can record the output of an SN P system:

\begin{itemize}
    \item The moments of time at which the output neuron $i_0$ sends a spike i.e. if the neuron $i_0$ releases spikes at the moments $q_1,q_2\dots$ then the output of $\Pi$ is the sequence $q_1,q_2\dots$.
    \item The interval between the moments at which the output neuron $i_0$ sends a spike i.e. if the neuron $i_0$ releases spikes at the moments $q_1,q_2\dots$ then the output of $\Pi$ is the sequence $q_2-q_1, q_3-q_2,\dots$
\end{itemize}

An SN P system is constructed using the principle of minimal determinism i.e. at a certain moment in time, either a firing or a forgetting rule is applied without being able to choose which of the two types of rules is applied  \cite{ionescu2006spiking}.

\subsection{Homomorphic encryption}

In this work, we use the ElGamal cryptosystem  \cite{elgamal1985public}. The security of the encryption scheme is based on the computational Diffie-Hellman assumption (CDH). Moreover, the scheme achieves semantic security based on the decisional Diffie-Hellman assumption (DDH) i.e. the scheme is randomized. Randomization implies that when encrypting the same message multiple times, each resulting ciphertext will be different. A consequence of this property is the fact that an attacker cannot distinguish two plaintexts by analyzing the corresponding ciphertext with non-negligible probability. The scheme works over a group $G$ of order $q$ with a generator $g$. We now proceed to the description of the  cryptosystem:

\begin{itemize}
    \item The key generation algorithm denoted by $\textbf{KeyGen}$ generates a key pair i.e. a private key and the corresponding public key. The algorithm takes the following steps:
    \begin{enumerate}
        \item Generate a random integer $x \in \{1,2,\dots q-1\}$.
        \item Compute $h:=g^x$.
        \item Output the public key $h$ and the corresponding private key $x$.
    \end{enumerate}
    \item The encryption algorithm denoted by $\textbf{Enc}^y_h$ encrypts a plaintext $m \in G$ using the public key $h$ and a random number $y \in \{1,2,\dots,q-1\}$. The algorithm performs the following steps:
    \begin{enumerate}
        \item Computes $s:=h^y$.
        \item Computes $c_1:=g^y$ and $c_2:=m \cdot s$.
        \item Outputs the ciphertext $c:=\left(c_1, c_2\right)$.
    \end{enumerate}
    \item The decryption algorithm denoted by $\textbf{Dec}_x$ takes as input a ciphertext $c=\left(c_1, c_2\right)$ and decrypts it under the private key $x$. The algorithm is composed of the following steps:
    \begin{enumerate}
        \item Computes $s:=c_1^x$.
        \item Computes $s^{-1}$, the inverse of $s$ in the group $G$.
        \item Computes the plaintext $m:=c_2 \cdot s^{-1}$.
        \item Outputs the plaintext $m$.
    \end{enumerate}
\end{itemize}
The proof of correctness is straightforward:
\begin{equation}
    c_2 \cdot s^{-1} = c_2 \cdot c_1^{-x} = m \cdot h^y \cdot g^{-xy} = m \cdot g^{-xy} \cdot g^{xy} = m 
\end{equation}

The encryption of a message $m \in G$ can be summarized by the following two equations:
\begin{equation}
\label{eq:-1}
c_1 = g^y
\end{equation}
\begin{equation}
\label{eq:-2}
c_2 = m \cdot h^y
\end{equation}
The scheme is homomorphic with respect to multiplication. Let $c=\left(c_1, c_2\right)$ and $c'=\left(c_1', c_2'\right)$ be the encryptions of two plaintexts $m$ and $m'$ under the same public key $h$ i.e. $c=\textbf{Enc}^y_h\left(m\right)$ and $c'=\textbf{Enc}^y_h\left(m'\right)$. We define the following two operations:
\begin{enumerate}
    \item Let $c \odot c' = \left(c_1 \cdot c_1', c_2 \cdot c_2'\right)$ be the multiplication of two ciphertexts. The result of this operation is another ciphertext that encrypts the sum between $m$ and $m'$:
    \begin{equation}
    \label{eq:1}
        c_1 \cdot c_1' = g^y \cdot g^{y'} = g^{y+y'}
    \end{equation}
    \begin{equation}
    \label{eq:2}
        c_2 \cdot c_2' = m \cdot s \cdot m' \cdot s' = 
        m \cdot m' \cdot h^y \cdot h^{y'} = m \cdot m' \cdot h^{y + y'}
    \end{equation}
    From \ref{eq:1} and \ref{eq:2} we can see that $c \odot c' = \left(c_1 \cdot c_1', c_2 \cdot c_2'\right)$ is a ciphertext that encrypts $m \cdot m'$ under the public key $h$.

    \item Let $c \otimes k=\left(c_1 , c_2 \cdot k \right)$ be the multiplication between a ciphertext and a constant $k$. The result of this operation is a ciphertext that encrypts the product between $m$ and $k$ as it can be seen from \ref{eq:3} and \ref{eq:4}:
    \begin{equation}
    \label{eq:3}
    c_1 = g^y
    \end{equation}
    \begin{equation}
    \label{eq:4}
    c_2 \cdot k = m \cdot s \cdot k = \left(m\cdot k\right) \cdot s = \left(m \cdot k\right) h^y
    \end{equation}
    \item When $c_1=c_1'$ i.e. $y=y'$, we can also define the addition between two ciphertexts as follows: $c \oplus c'=\left(c_1, c_2 + c_2'\right)$. The result of this operation is a ciphertext that encrypts the sum between $m$ and $m'$:
    \begin{equation}
    \label{eq:5}
    c_1 = g^y
    \end{equation}
    \begin{equation}
    \label{eq:6}
    c_2 + c_2' = m \cdot s + m' \cdot s = \left(m+m'\right) \cdot s = \left(m+m'\right) \cdot h^y
    \end{equation}
    
    It is important to notice that when $y=y'$ the scheme is no longer semantically secure. Although an attacker who observes the two ciphertexts $c$ and $c'$ could not recover any of the plaintexts, it could determine additional information about them e.g. whether they are different. In many scenarios, this is not acceptable but in this work, we will use the $\oplus$ operation. All the messages encrypted with the same random $y$ are not critical i.e. the impact of the lack of semantic security does not affect the security of the protocol in which the encryption scheme is used.

\end{enumerate}

\section{SN P system to compute linear functions}
In this section, we describe an SN P system that computes functions of the form $t_1k + t_2$ over natural numbers. 

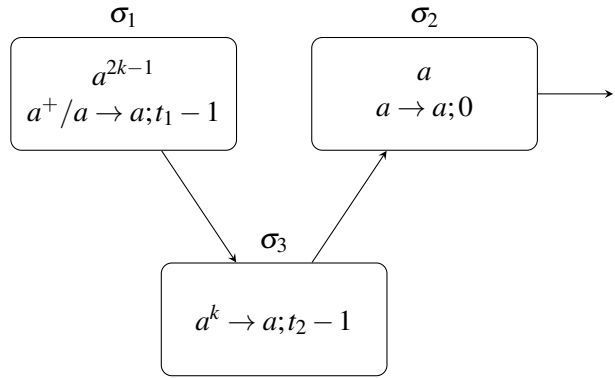
\begin{figure}[!htb]
\centering
\begin{tikzpicture}[>=stealth, rounded corners=10pt]
    \node[draw, rounded corners, minimum width=3cm, minimum height=1.5cm, align=center, label=above:{$\sigma_1$}] (oval1) at (4, 0) {$a^{2k-1}$\\$a^+/a \rightarrow a; t_1-1$};

    \node[draw, rounded corners, minimum width=3cm, minimum height=1.5cm, align=center, label=above:{$\sigma_2$}] (oval2) at (8, 0) {$a$\\$a \rightarrow a;0$};

    \node[draw, rounded corners, minimum width=3cm, minimum height=1.5cm, align=center, label=above:{$\sigma_3$}] (oval3) at (6, -3) {$a^k \rightarrow a;t_2-1$};

    \draw[->] (oval1) -- (oval3);
    \draw[->] (oval3) -- (oval2);
    \draw[->] (oval2.east) -- ++(1cm, 0);
\end{tikzpicture}
\caption{ SN P system to compute linear functions} 
\label{fig:1}
\end{figure}

Let $\Pi_{Add}\left(t_1, t_2, k\right)=\{\{a\}, \sigma_1, \sigma_2, \sigma_3, syn_{add}, \sigma_2\}$ be the SN P system that computes the linear function $t_1k+t_2$,\ $t_1, t_2 \in \mathbf{N}$ with the following components:
\begin{itemize}
    \item The alphabet is made from a single symbol $\{a\}$ that denotes a spike.
    \item There are three neurons: $\sigma_1, \sigma_2$ and $\sigma_3$ with the following firing rules:
    \begin{enumerate}
        \item For $\sigma_1$ the firing rule is $a^+/a \rightarrow a; t_1-1$.
        \item For $\sigma_2$ the firing rule is $a \rightarrow a;0$.
        \item For $\sigma_3$ the firing rule is $a^k \rightarrow a; t_2 -1 $.
    \end{enumerate}
    \item The set of synapses $syn_{add}$ is the set $\{\left(1,3\right), \left(3,2\right)\}$.
    \item The output neuron is $\sigma_2$.
\end{itemize}

Initially $\sigma_1$ has $2k-1$ spikes, $\sigma_2$ has $1$ spike and $\sigma_3$ has no spikes. Since $\sigma_2$ has no refractory time and one spike, it will release it in the first step of the computation. During the first step, $\sigma_1$ will be firing. Since its refractory period is $t_1 -1$ during the time steps $2, 3, \dots t_1-1$ the neuron will be closed i.e. it will not receive or send any spikes. In step $t_1$ the neuron will send one spike to $\sigma_3$ and fire again. Thus $\sigma_1$ fires every $t_1$ steps consuming one spike and sending one spike to $\sigma_3$. $\sigma_3$ will fire when it acumulates $k$ spikes from neuron $\sigma_1$. Since $\sigma_1$ fires one spikes every $t_1$ steps, at time step, $k\cdot t_1$ $\sigma_3$ will receive the $k^{th}$ spike. At the moment $t_1 \cdot k +1$, $\sigma_3$ will fire. The refractory period of this neuron is $t_2 - 1$ thus at the steps $t_1 \cdot k +2, t_1 \cdot k +3, \dots, t_1 \cdot k +t_2 -1$ it will be closed and it will release one spike to $\sigma_2$ at $t_1 \cdot k +t_2$. Since $\sigma_3$ has no refractory period, it will release the spike at the moment $t_1 \cdot k + t_2 +1$. At this point, the number of spikes left in $\sigma_1$ is $k-1$ because it already sent $k$ spikes to neuron $\sigma_3$ and the initial number of spikes was $2k-1$. The neuron will continue to send spikes to $\sigma_ 3$ at the appropriate time steps until the spikes run out. The neuron $\sigma_3$ will never fire again since it can no longer accumulate $k$ spikes. $\sigma_2$ will never fire again either since it will no longer receive the spike from $\sigma_3$. Thus, after $\sigma_1$ exhausts all the spikes, the computation will stop. There are two moments when the output neuron $\sigma_2$ fires: $1$ and $t_1 \cdot k + t_2 +1$. Thus, the result of the computation i.e. the difference between the time points at which the output neuron fires, is $t_1 \cdot k + t_2$. The $\Pi_{Add}\left(t_1, t_2, k\right)$ system is depicted in Figure \ref{fig:1}.

\section{Privacy-preserving computations in SN P systems}

In this section, we describe our protocol which enables the running of an SN P system to compute linear functions in a privacy-preserving way. We also make some remarks regarding security.

\subsection{The protocol}

There are two actors in the protocol:
\begin{enumerate}
    \item The Server: it can instantiate and run an SN P system of the form $\Pi_{Add}\left(t_1, t_2, k\right)$ for any integers $t_1, t_2$ and $k$.
    \item The Client: it wants to evaluate the linear function $t_1 \cdot k + t_2$ using the system hosted by the Server without revealing any of inputs $t_1, t_2$ or $k$.
\end{enumerate}

The protocol uses the holomorphic properties of the ElGamal cryptosystem to allow the client to use the server without revealing the inputs of the SN P system. There are $7$ steps:

\begin{enumerate}
    \item The Client will use the $\textbf{KeyGen}$ algorithm to generate a key pair: $h$ and $x$. 
    \item The Client will use the encryption algorithm $\textbf{Enc}^y_h$ to encrypt $t_1, t_2$ and $k$. We denote by $c^{t_1}=\left(c^{t_1}_1, c^{t_1}_2\right), c^{t_2}=\left(c^{t_2}_1, c^{t_2}_2\right)$ and $c^k = \left(c^{t_k}_1, c^{k}_2\right)$ the encryptions of $t_1, t_2$ and $k$:
    \begin{itemize}
        \item $c^{t_1}=\left(c^{t_1}_1, c^{t_1}_2\right) = \textbf{Enc}^{y_1}_h\left(t_1\right)$
        \item $c^k = \left(c^{k}_1, c^{k}_2\right) = \textbf{Enc}^{y_2}_h\left(k\right)$
        \item $c^{t_2}=\left(c^{t_2}_1, c^{t_2}_2\right) = \textbf{Enc}^{y_1 + y_2}_h\left(t_2\right)$
    \end{itemize}
    \item The Client will store locally $c^{t_1}_1, c^{t_2}_1$ and $c^{k}_1$ and send to the server $c^{t_1}_2, c^{t_2}_2$ and $c^{k}_2$. 
    \item The Server will instantiate and run an SN P system of the form $\Pi_{Add}\left(c^{t_1}_2, c^{t_2}_2, c^{k}_2\right)$. 
    \item After the computation stops, the Server will return to the client the result of the computation i.e. $c_2 = c^{t_1}_2 \cdot c^{k}_2 + c^{t_2}_2$.
    \item The Client will compose a new ciphertext $c=\left(c_1, c_2\right)$ where $c_1 = c^{t_1}_1 \cdot c^{k}_1$.
    \item The Client will decrypt the ciphertext $c$ using the algorithm $\textbf{Dec}_x$  and retrived the result of the computation i.e. $t_1 \cdot k + t_2$.

\end{enumerate}

We now prove that the ciphertext computed in step $6$ of the protocol is a valid ElGamal ciphertext that correctly decrypts to the final result of the computation i.e. $t_1 \cdot k + t_2$.
Let $c'$ be the following ciphertext:
\begin{equation}
    c'=\left(c_1, c^{t_1}_2 \cdot c^{k}_2\right)
\end{equation}
Since $c_1=c^{t_1}_1 \cdot c^{k}_1$ we can write $c'$ as:
\begin{equation}
    c'=\left(c^{t_1}_1 \cdot c^{k}_1, c^{t_1}_2 \cdot c^{k}_2\right)
\end{equation}
The ciphertext $c'$ represents the multiplication between the ciphertexts $c^{t_1}$ and $c^k$:
\begin{equation}
    c'= c^{t_1} \odot c^k
\end{equation}

Since $c$ and $c'$ use the same randomess i.e. $c_1$, we can write $c$ as the sum between the ciphertext $c'$ and  $c^{t_2}$:
\begin{equation}
    c = c' \oplus c^{t_2}
\end{equation}
In conclusion, we can express $c$ as a composition of valid ElGamal ciphertexts which is also a valid ElGamal ciphertext:
\begin{equation}
    c = \left(c^{t_1} \odot c^k\right) \oplus c^{t_2}
\end{equation}

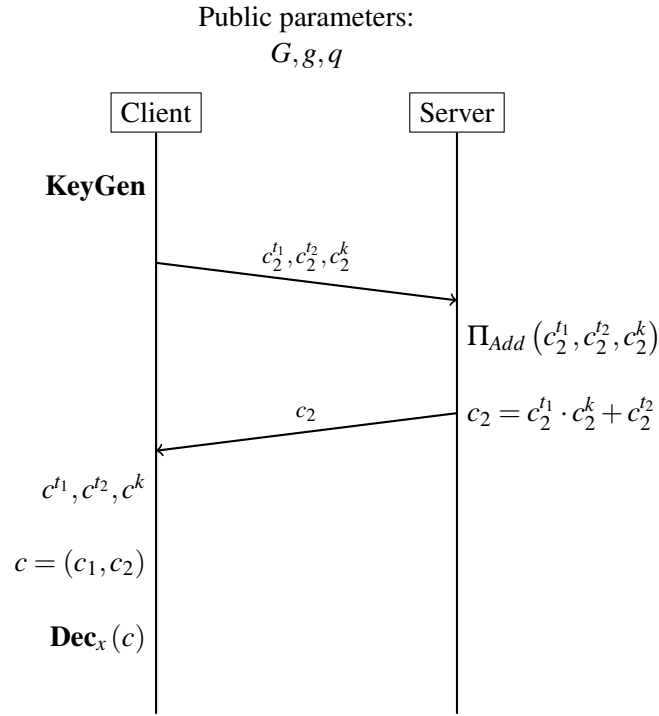
\begin{figure}[!htb]
\centering
\begin{tikzpicture}
   \node[draw=none,fill=none,align=center] (public) at (0,1) {Public parameters:\\$G, g, q$};
  \node[draw] (Prover) at (-2,0) {Client}; 
  \draw[thick] (Prover) -- ++(0, -8);
    
  \node[draw=none,fill=none,anchor=east] (asecret) at ($(Prover) + (0,-1)$) {$\textbf{KeyGen}$};
  \node[draw=none,fill=none,anchor=east] (asecret) at ($(Prover) + (0,-5)$) {$c^{t_1}, c^{t_2}, c^{k}$};
  \node[draw=none,fill=none,anchor=east] (asecret) at ($(Prover) + (0,-6)$) {$c=\left(c_1, c_2\right)$};
  \node[draw=none,fill=none,anchor=east] (asecret) at ($(Prover) + (0,-7)$) {$\textbf{Dec}_x\left(c\right)$};

  \node[draw] (Verifier) at (2,0) {Server}; 
  \draw[thick] (Verifier) -- ++(0, -8);
   
  \node[draw=none,fill=none,anchor=west] (bsecret) at ($(Verifier) + (0,-3)$) {$\Pi_{Add}\left(c^{t_1}_2, c^{t_2}_2, c^{k}_2\right)$};

   \node[draw=none,fill=none,anchor=west] (bsecret) at ($(Verifier) + (0,-4)$) {$c_2 = c^{t_1}_2 \cdot c^{k}_2 + c^{t_2}_2$};

  \draw[->,thick] ($(Prover)+(0,-2)$) -- ($(Verifier)+(0,-2.5)$) node [pos=0.5,above,font=\footnotesize] {${c^{t_1}_2, c^{t_2}_2, c^{k}_2}$};

   \draw[->,thick] ($(Verifier)+(0,-4)$) -- ($(Prover)+(0,-4.5)$) node [pos=0.5,above,font=\footnotesize] {$c_2$};

\end{tikzpicture}
\caption{ Privacy-preserving linear function computation using SN P systems} 
\label{fig:2}
\end{figure}

The ciphertext $\left(c^{t_1} \odot c^k\right)$ represents the encryption of $t_1 \cdot k$. When we add this ciphertext with $c^{t_2}$ using the $\oplus$ operation, the resulting ciphertext will be the encryption of $t_1 \cdot k + t_2$ which is the result of the computation performed over plaintext data. Figure \ref{fig:2} depicts our protocol.

Since the SN P system works over natural numbers, we can use $G=\mathbf{Z}_q$ for a large prime $q$.

\subsection{Security considerations}
We analyze our protocol in the honest-but-curious security model \cite{paverd2014modelling}. In this model, we assume that the adversary is the Server. There are two properties of the adversary:

\begin{enumerate}
    \item Curious: the Server will try to find information about the underlying plaintexts. In our case, these are the parameters of the SN P system: $t_1, t_2$, and $k$.
    \item Honest: the Server will respect the protocol and it will complete every step. It will not modify in any way the messages received or sent to the Client. 
\end{enumerate}

The underlying encryption scheme i.e. the ElGamal cryptosystem is semantically secure given the DDH assumption as long as each ciphertext is generated using different randomness \cite{goldreich2004foundations}. The security of the scheme can be illustrated using the following game. We suppose that the attacker runs in probabilistic polynomial time and has access to an encryption oracle that receives as input a plaintext and returns the corresponding ciphertext:
\begin{itemize}
    \item The attacker chooses as many plaintexts as it wants and encrypts them using the oracle. For each ciphertext, the attacker knows the corresponding plaintext. 
    \item The attacker chooses two plaintext $m_0$ and $m_1$ and send them to the oracle.
    \item The oracle will generate a random bit $b$ and return the encryption of $m_b$.
    \item The attacker outputs the bit $b$.
\end{itemize}
The scheme is secure as long as the attacker cannot output the correct bit with non-negligible probability. 

Given the fact that the Client encrypts each parameter of the SN P system using the ElGamal cryptosystem with different randomness, the Server cannot learn any information about them with non-negligible probability. 

\section{Conclusions and further directions of research}

In this paper, we presented a protocol for performing privacy-preserving computation over SN P systems. There are two actors involved: the client and the server. The server hosts an SN P system that computes linear function over natural numbers i.e. $t_1k +t_2$. The client uses the protocol to retrieve the result of the computation without revealing $t_1, t_2$ or $k$ and without the server knowing the result of the calculation. We presented an SN P system that computes any linear function over natural numbers and also some security considerations about our protocol which is based on the ElGamal cryptosystem. 

There are several directions of research. The first one is to give formal proof of the security of the protocol. Although the protocol is secure at first sight, we must prove it by reducing the security of it to the security of the underlying cryptosystem. The second direction of research is to enable complex computation on SN P systems in a privacy-preserving way. The third direction is to use dedicated cryptosystems e.g. fully homomorphic encryption schemes which are created to enable privacy-preserving computations  \cite{martins2017survey}. The challenge here is to map the operations performed over encrypted data to the operations performed by the SN P system. We should also consider using other privacy-enhancing technologies e.g. secure multi-party computation or differential privacy to enable private computations over SN P systems. Another direction of research is to implement the protocol and perform the appropriate benchmarking to study its efficiency and communication overhead in practice.

\subsubsection{Acknowledgements} This research was supported by the European Regional Development Fund, Competitiveness Operational Program 2014-2020 through project IDBC (code SMIS 2014+: 121512).

\bibliographystyle{eptcs}
\bibliography{generic}

\begin{thebibliography}{10}
\providecommand{\bibitemdeclare}[2]{}
\providecommand{\surnamestart}{}
\providecommand{\surnameend}{}
\providecommand{\urlprefix}{Available at }
\providecommand{\url}[1]{\texttt{#1}}
\providecommand{\href}[2]{\texttt{#2}}
\providecommand{\urlalt}[2]{\href{#1}{#2}}
\providecommand{\doi}[1]{doi:\urlalt{https://doi.org/#1}{#1}}
\providecommand{\eprint}[1]{arXiv:\urlalt{https://arxiv.org/abs/#1}{#1}}
\providecommand{\bibinfo}[2]{#2}

\bibitemdeclare{article}{arteta2021vivo}
\bibitem{arteta2021vivo}
\bibinfo{author}{Alberto \surnamestart Arteta~Albert\surnameend},
  \bibinfo{author}{Ernesto \surnamestart D{\'\i}az-Flores\surnameend},
  \bibinfo{author}{Luis Fernando de~Mingo \surnamestart L{\'o}pez\surnameend}
  \& \bibinfo{author}{Nuria \surnamestart G{\'o}mez~Blas\surnameend}
  (\bibinfo{year}{2021}): \emph{\bibinfo{title}{An in vivo proposal of cell
  computing inspired by membrane computing}}.
\newblock {\slshape \bibinfo{journal}{Processes}}
  \bibinfo{volume}{9}(\bibinfo{number}{3}), p. \bibinfo{pages}{511},
  \doi{10.3390/pr9030511}.

\bibitemdeclare{article}{chen2021survey}
\bibitem{chen2021survey}
\bibinfo{author}{Yunhui \surnamestart Chen\surnameend}, \bibinfo{author}{Ying
  \surnamestart Chen\surnameend}, \bibinfo{author}{Gexiang \surnamestart
  Zhang\surnameend}, \bibinfo{author}{Prithwineel \surnamestart
  Paul\surnameend}, \bibinfo{author}{Tianbao \surnamestart Wu\surnameend},
  \bibinfo{author}{Xihai \surnamestart Zhang\surnameend},
  \bibinfo{author}{Haina \surnamestart Rong\surnameend} \&
  \bibinfo{author}{Xiaomin \surnamestart Ma\surnameend} (\bibinfo{year}{2021}):
  \emph{\bibinfo{title}{{A Survey of Learning Spiking Neural P Systems and A
  Novel Instance}}}.
\newblock {\slshape \bibinfo{journal}{International Journal of Unconventional
  Computing}} \bibinfo{volume}{16}.

\bibitemdeclare{incollection}{csuhaj2022membrane}
\bibitem{csuhaj2022membrane}
\bibinfo{author}{Erzs{\'e}bet \surnamestart Csuhaj-Varj{\'u}\surnameend},
  \bibinfo{author}{Marian \surnamestart Gheorghe\surnameend},
  \bibinfo{author}{Alberto \surnamestart Leporati\surnameend},
  \bibinfo{author}{Miguel~{\'A}ngel \surnamestart Mart{\'\i}nez-del
  Amor\surnameend}, \bibinfo{author}{Linqiang \surnamestart Pan\surnameend},
  \bibinfo{author}{Prithwineel \surnamestart Paul\surnameend},
  \bibinfo{author}{Andrei \surnamestart P{\u{a}}un\surnameend},
  \bibinfo{author}{Ignacio \surnamestart P{\'e}rez-Hurtado\surnameend},
  \bibinfo{author}{Mario~J \surnamestart P{\'e}rez-Jim{\'e}nez\surnameend},
  \bibinfo{author}{Bosheng \surnamestart Song\surnameend} et~al.
  (\bibinfo{year}{2022}): \emph{\bibinfo{title}{{Membrane computing concepts,
  theoretical developments and applications}}}.
\newblock In: {\slshape \bibinfo{booktitle}{Handbook of Unconventional
  Computing: VOLUME 1: Theory}}, \bibinfo{publisher}{World Scientific}, pp.
  \bibinfo{pages}{261--339}, \doi{10.1142/9789811235726_0008}.

\bibitemdeclare{article}{diaz2019membrane}
\bibitem{diaz2019membrane}
\bibinfo{author}{Daniel \surnamestart D{\'\i}az-Pernil\surnameend},
  \bibinfo{author}{Miguel~A \surnamestart Guti{\'e}rrez-Naranjo\surnameend} \&
  \bibinfo{author}{Hong \surnamestart Peng\surnameend} (\bibinfo{year}{2019}):
  \emph{\bibinfo{title}{Membrane computing and image processing: a short
  survey}}.
\newblock {\slshape \bibinfo{journal}{Journal of Membrane Computing}}
  \bibinfo{volume}{1}, pp. \bibinfo{pages}{58--73},
  \doi{10.1007/s41965-018-00002-x}.

\bibitemdeclare{article}{diaz2013parallel}
\bibitem{diaz2013parallel}
\bibinfo{author}{Daniel \surnamestart D{\'\i}az-Pernil\surnameend},
  \bibinfo{author}{Francisco \surnamestart Pena-Cantillana\surnameend} \&
  \bibinfo{author}{Miguel~A \surnamestart Guti{\'e}rrez-Naranjo\surnameend}
  (\bibinfo{year}{2013}): \emph{\bibinfo{title}{{A parallel algorithm for
  skeletonizing images by using spiking neural P systems}}}.
\newblock {\slshape \bibinfo{journal}{Neurocomputing}} \bibinfo{volume}{115},
  pp. \bibinfo{pages}{81--91}, \doi{10.1016/j.neucom.2012.12.032}.

\bibitemdeclare{article}{elgamal1985public}
\bibitem{elgamal1985public}
\bibinfo{author}{Taher \surnamestart ElGamal\surnameend}
  (\bibinfo{year}{1985}): \emph{\bibinfo{title}{A public key cryptosystem and a
  signature scheme based on discrete logarithms}}.
\newblock {\slshape \bibinfo{journal}{IEEE transactions on information theory}}
  \bibinfo{volume}{31}(\bibinfo{number}{4}), pp. \bibinfo{pages}{469--472},
  \doi{10.1109/TIT.1985.1057074}.

\bibitemdeclare{article}{fan2021implementation}
\bibitem{fan2021implementation}
\bibinfo{author}{Songhai \surnamestart Fan\surnameend}, \bibinfo{author}{Yiyu
  \surnamestart Gong\surnameend}, \bibinfo{author}{Gexiang \surnamestart
  Zhang\surnameend}, \bibinfo{author}{Yun \surnamestart Xiao\surnameend},
  \bibinfo{author}{Haina \surnamestart Rong\surnameend},
  \bibinfo{author}{Prithwineel \surnamestart Paul\surnameend},
  \bibinfo{author}{Xiaomin \surnamestart Ma\surnameend}, \bibinfo{author}{Han
  \surnamestart Huang\surnameend} \& \bibinfo{author}{Marian \surnamestart
  Gheorghe\surnameend} (\bibinfo{year}{2021}):
  \emph{\bibinfo{title}{{Implementation of Kernel P Systems in CUDA for Solving
  NP-hard Problems}}}.
\newblock {\slshape \bibinfo{journal}{International Journal of Unconventional
  Computing}} \bibinfo{volume}{16}.

\bibitemdeclare{article}{ganbaatar2021implementation}
\bibitem{ganbaatar2021implementation}
\bibinfo{author}{Ganbat \surnamestart Ganbaatar\surnameend},
  \bibinfo{author}{Dugar \surnamestart Nyamdorj\surnameend},
  \bibinfo{author}{Gordon \surnamestart Cichon\surnameend} \&
  \bibinfo{author}{Tseren-Onolt \surnamestart Ishdorj\surnameend}
  (\bibinfo{year}{2021}): \emph{\bibinfo{title}{{Implementation of RSA
  cryptographic algorithm using SN P systems based on HP/LP neurons}}}.
\newblock {\slshape \bibinfo{journal}{Journal of Membrane Computing}}
  \bibinfo{volume}{3}, pp. \bibinfo{pages}{22--34},
  \doi{10.1007/s41965-021-00073-3}.

\bibitemdeclare{book}{goldreich2004foundations}
\bibitem{goldreich2004foundations}
\bibinfo{author}{Oded \surnamestart Goldreich\surnameend}
  (\bibinfo{year}{2004}): \emph{\bibinfo{title}{Foundations of Cryptography,
  Volume 2}}.
\newblock \bibinfo{publisher}{Cambridge university press Cambridge},
  \doi{10.1017/CBO9780511721656}.

\bibitemdeclare{inproceedings}{guo2016family}
\bibitem{guo2016family}
\bibinfo{author}{Ping \surnamestart Guo\surnameend} \& \bibinfo{author}{Wei
  \surnamestart Xu\surnameend} (\bibinfo{year}{2016}): \emph{\bibinfo{title}{{A
  family P system of realizing RSA algorithm}}}.
\newblock In: {\slshape \bibinfo{booktitle}{Bio-inspired Computing--Theories
  and Applications: 11th International Conference, BIC-TA 2016, Xi'an, China,
  October 28-30, 2016, Revised Selected Papers, Part I 11}},
  \bibinfo{organization}{Springer}, pp. \bibinfo{pages}{155--167},
  \doi{10.1007/978-981-10-3611-8_16}.

\bibitemdeclare{article}{guo2017implementation}
\bibitem{guo2017implementation}
\bibinfo{author}{Ping \surnamestart Guo\surnameend} \& \bibinfo{author}{Wei
  \surnamestart Xu\surnameend} (\bibinfo{year}{2017}):
  \emph{\bibinfo{title}{{Implementation of RSA algorithm based on P system}}}.
\newblock {\slshape \bibinfo{journal}{Journal of Computational and Theoretical
  Nanoscience}} \bibinfo{volume}{14}(\bibinfo{number}{9}), pp.
  \bibinfo{pages}{4227--4235}, \doi{10.1166/jctn.2017.6723}.

\bibitemdeclare{article}{ionescu2006spiking}
\bibitem{ionescu2006spiking}
\bibinfo{author}{Mihai \surnamestart Ionescu\surnameend},
  \bibinfo{author}{Gheorghe \surnamestart P{\u{a}}un\surnameend} \&
  \bibinfo{author}{Takashi \surnamestart Yokomori\surnameend}
  (\bibinfo{year}{2006}): \emph{\bibinfo{title}{Spiking neural P systems}}.
\newblock {\slshape \bibinfo{journal}{Fundamenta informaticae}}
  \bibinfo{volume}{71}(\bibinfo{number}{2-3}), pp. \bibinfo{pages}{279--308}.

\bibitemdeclare{article}{martins2017survey}
\bibitem{martins2017survey}
\bibinfo{author}{Paulo \surnamestart Martins\surnameend},
  \bibinfo{author}{Leonel \surnamestart Sousa\surnameend} \&
  \bibinfo{author}{Artur \surnamestart Mariano\surnameend}
  (\bibinfo{year}{2017}): \emph{\bibinfo{title}{{A survey on fully homomorphic
  encryption: An engineering perspective}}}.
\newblock {\slshape \bibinfo{journal}{ACM Computing Surveys (CSUR)}}
  \bibinfo{volume}{50}(\bibinfo{number}{6}), pp. \bibinfo{pages}{1--33},
  \doi{10.1145/3124441}.

\bibitemdeclare{article}{pan2017spiking}
\bibitem{pan2017spiking}
\bibinfo{author}{Linqiang \surnamestart Pan\surnameend},
  \bibinfo{author}{Gheorghe \surnamestart P{\u{a}}un\surnameend},
  \bibinfo{author}{Gexiang \surnamestart Zhang\surnameend} \&
  \bibinfo{author}{Ferrante \surnamestart Neri\surnameend}
  (\bibinfo{year}{2017}): \emph{\bibinfo{title}{{Spiking neural P systems with
  communication on request}}}.
\newblock {\slshape \bibinfo{journal}{International journal of neural systems}}
  \bibinfo{volume}{27}(\bibinfo{number}{08}), p. \bibinfo{pages}{1750042},
  \doi{10.1142/S0129065717500423}.

\bibitemdeclare{article}{pan2010computational}
\bibitem{pan2010computational}
\bibinfo{author}{Linqiang \surnamestart Pan\surnameend} \&
  \bibinfo{author}{Mario~J \surnamestart P{\'e}rez-Jim{\'e}nez\surnameend}
  (\bibinfo{year}{2010}): \emph{\bibinfo{title}{{Computational complexity of
  tissue-like P systems}}}.
\newblock {\slshape \bibinfo{journal}{Journal of Complexity}}
  \bibinfo{volume}{26}(\bibinfo{number}{3}), pp. \bibinfo{pages}{296--315},
  \doi{10.1016/j.jco.2010.03.001}.

\bibitemdeclare{article}{pan2012spiking}
\bibitem{pan2012spiking}
\bibinfo{author}{Linqiang \surnamestart Pan\surnameend}, \bibinfo{author}{Jun
  \surnamestart Wang\surnameend} \& \bibinfo{author}{Hendrik~Jan \surnamestart
  Hoogeboom\surnameend} (\bibinfo{year}{2012}): \emph{\bibinfo{title}{{Spiking
  neural P systems with astrocytes}}}.
\newblock {\slshape \bibinfo{journal}{Neural Computation}}
  \bibinfo{volume}{24}(\bibinfo{number}{3}), pp. \bibinfo{pages}{805--825},
  \doi{10.1162/NECO_a_00238}.

\bibitemdeclare{article}{puaun2000computing}
\bibitem{puaun2000computing}
\bibinfo{author}{Gheorghe \surnamestart P{\u{a}}un\surnameend}
  (\bibinfo{year}{2000}): \emph{\bibinfo{title}{Computing with membranes}}.
\newblock {\slshape \bibinfo{journal}{Journal of Computer and System Sciences}}
  \bibinfo{volume}{61}(\bibinfo{number}{1}), pp. \bibinfo{pages}{108--143},
  \doi{10.1006/jcss.1999.1693}.

\bibitemdeclare{article}{paverd2014modelling}
\bibitem{paverd2014modelling}
\bibinfo{author}{Andrew \surnamestart Paverd\surnameend},
  \bibinfo{author}{Andrew \surnamestart Martin\surnameend} \&
  \bibinfo{author}{Ian \surnamestart Brown\surnameend} (\bibinfo{year}{2014}):
  \emph{\bibinfo{title}{Modelling and automatically analysing privacy
  properties for honest-but-curious adversaries}}.
\newblock {\slshape \bibinfo{journal}{Tech. Rep}}.

\bibitemdeclare{article}{peng2015novel}
\bibitem{peng2015novel}
\bibinfo{author}{Hong \surnamestart Peng\surnameend}, \bibinfo{author}{Xiaohui
  \surnamestart Luo\surnameend}, \bibinfo{author}{Zhisheng \surnamestart
  Gao\surnameend}, \bibinfo{author}{Jun \surnamestart Wang\surnameend},
  \bibinfo{author}{Zheng \surnamestart Pei\surnameend} et~al.
  (\bibinfo{year}{2015}): \emph{\bibinfo{title}{A novel clustering algorithm
  inspired by membrane computing}}.
\newblock {\slshape \bibinfo{journal}{The Scientific World Journal}}
  \bibinfo{volume}{2015}, \doi{10.1155/2015/929471}.

\bibitemdeclare{article}{peng2015unsupervised}
\bibitem{peng2015unsupervised}
\bibinfo{author}{Hong \surnamestart Peng\surnameend}, \bibinfo{author}{Jun
  \surnamestart Wang\surnameend}, \bibinfo{author}{Mario~J \surnamestart
  P{\'e}rez-Jim{\'e}nez\surnameend} \& \bibinfo{author}{Agust{\'\i}n
  \surnamestart Riscos-N{\'u}{\~n}ez\surnameend} (\bibinfo{year}{2015}):
  \emph{\bibinfo{title}{An unsupervised learning algorithm for membrane
  computing}}.
\newblock {\slshape \bibinfo{journal}{Information Sciences}}
  \bibinfo{volume}{304}, pp. \bibinfo{pages}{80--91},
  \doi{10.1016/j.ins.2015.01.019}.

\bibitemdeclare{article}{peng2015automatic}
\bibitem{peng2015automatic}
\bibinfo{author}{Hong \surnamestart Peng\surnameend}, \bibinfo{author}{Jun
  \surnamestart Wang\surnameend}, \bibinfo{author}{Peng \surnamestart
  Shi\surnameend}, \bibinfo{author}{Agust{\'\i}n \surnamestart
  Riscos-N{\'u}{\~n}ez\surnameend} \& \bibinfo{author}{Mario~J \surnamestart
  P{\'e}rez-Jim{\'e}nez\surnameend} (\bibinfo{year}{2015}):
  \emph{\bibinfo{title}{An automatic clustering algorithm inspired by membrane
  computing}}.
\newblock {\slshape \bibinfo{journal}{Pattern Recognition Letters}}
  \bibinfo{volume}{68}, pp. \bibinfo{pages}{34--40},
  \doi{10.1016/j.patrec.2015.08.008}.

\bibitemdeclare{article}{plesa2022key}
\bibitem{plesa2022key}
\bibinfo{author}{Mihail-Iulian \surnamestart Plesa\surnameend},
  \bibinfo{author}{Marian \surnamestart Gheoghe\surnameend},
  \bibinfo{author}{Florentin \surnamestart Ipate\surnameend} \&
  \bibinfo{author}{Gexiang \surnamestart Zhang\surnameend}
  (\bibinfo{year}{2022}): \emph{\bibinfo{title}{{A key agreement protocol based
  on spiking neural P systems with anti-spikes}}}.
\newblock {\slshape \bibinfo{journal}{Journal of Membrane Computing}}
  \bibinfo{volume}{4}(\bibinfo{number}{4}), pp. \bibinfo{pages}{341--351},
  \doi{10.1007/s41965-022-00110-9}.

\bibitemdeclare{article}{song2017spiking}
\bibitem{song2017spiking}
\bibinfo{author}{Tao \surnamestart Song\surnameend}, \bibinfo{author}{Alfonso
  \surnamestart Rodr{\'\i}guez-Pat{\'o}n\surnameend}, \bibinfo{author}{Pan
  \surnamestart Zheng\surnameend} \& \bibinfo{author}{Xiangxiang \surnamestart
  Zeng\surnameend} (\bibinfo{year}{2017}): \emph{\bibinfo{title}{{Spiking
  neural P systems with colored spikes}}}.
\newblock {\slshape \bibinfo{journal}{IEEE Transactions on Cognitive and
  Developmental Systems}} \bibinfo{volume}{10}(\bibinfo{number}{4}), pp.
  \bibinfo{pages}{1106--1115}, \doi{10.1109/TCDS.2017.2785332}.

\bibitemdeclare{article}{vasile2023breaking}
\bibitem{vasile2023breaking}
\bibinfo{author}{R{\u{a}}zvan \surnamestart Vasile\surnameend},
  \bibinfo{author}{Marian \surnamestart Gheorghe\surnameend} \&
  \bibinfo{author}{Ionuț~Mihai \surnamestart Niculescu\surnameend}
  (\bibinfo{year}{2023}): \emph{\bibinfo{title}{{Breaking RSA Encryption
  Protocol with Kernel P Systems}}}.
\newblock \doi{10.21203/rs.3.rs-2684530/v1}.

\bibitemdeclare{article}{wu2017spiking}
\bibitem{wu2017spiking}
\bibinfo{author}{Tingfang \surnamestart Wu\surnameend}, \bibinfo{author}{Andrei
  \surnamestart P{\u{a}}un\surnameend}, \bibinfo{author}{Zhiqiang \surnamestart
  Zhang\surnameend} \& \bibinfo{author}{Linqiang \surnamestart Pan\surnameend}
  (\bibinfo{year}{2017}): \emph{\bibinfo{title}{{Spiking neural P systems with
  polarizations}}}.
\newblock {\slshape \bibinfo{journal}{IEEE transactions on neural networks and
  learning systems}} \bibinfo{volume}{29}(\bibinfo{number}{8}), pp.
  \bibinfo{pages}{3349--3360}, \doi{10.1109/TNNLS.2017.2726119}.

\bibitemdeclare{inproceedings}{yahya2016image}
\bibitem{yahya2016image}
\bibinfo{author}{Rafaa~I \surnamestart Yahya\surnameend},
  \bibinfo{author}{Siti~Mariyam \surnamestart Shamsuddin\surnameend},
  \bibinfo{author}{Salah~I \surnamestart Yahya\surnameend},
  \bibinfo{author}{Shafatnnur \surnamestart Hasan\surnameend},
  \bibinfo{author}{Bisan \surnamestart Al-Salibi\surnameend} \&
  \bibinfo{author}{Ghada \surnamestart Al-Khafaji\surnameend}
  (\bibinfo{year}{2016}): \emph{\bibinfo{title}{Image segmentation using
  membrane computing: a literature survey}}.
\newblock In: {\slshape \bibinfo{booktitle}{Bio-inspired Computing--Theories
  and Applications: 11th International Conference, BIC-TA 2016, Xi'an, China,
  October 28-30, 2016, Revised Selected Papers, Part I 11}},
  \bibinfo{organization}{Springer}, pp. \bibinfo{pages}{314--335},
  \doi{10.1007/978-981-10-3611-8_26}.

\bibitemdeclare{inproceedings}{zandron2001solving}
\bibitem{zandron2001solving}
\bibinfo{author}{Claudio \surnamestart Zandron\surnameend},
  \bibinfo{author}{Claudio \surnamestart Ferretti\surnameend} \&
  \bibinfo{author}{Giancarlo \surnamestart Mauri\surnameend}
  (\bibinfo{year}{2001}): \emph{\bibinfo{title}{{Solving NP-complete problems
  using P systems with active membranes}}}.
\newblock In: {\slshape \bibinfo{booktitle}{Unconventional Models of
  Computation, UMC’2K: Proceedings of the Second International Conference on
  Unconventional Models of Computation,(UMC’2K)}},
  \bibinfo{organization}{Springer}, pp. \bibinfo{pages}{289--301},
  \doi{10.1007/978-1-4471-0313-4_21}.

\bibitemdeclare{article}{zhang2008quantum}
\bibitem{zhang2008quantum}
\bibinfo{author}{Ge-Xiang \surnamestart Zhang\surnameend},
  \bibinfo{author}{Marian \surnamestart Gheorghe\surnameend} \&
  \bibinfo{author}{Chao-Zhong \surnamestart Wu\surnameend}
  (\bibinfo{year}{2008}): \emph{\bibinfo{title}{A quantum-inspired evolutionary
  algorithm based on P systems for knapsack problem}}.
\newblock {\slshape \bibinfo{journal}{Fundamenta Informaticae}}
  \bibinfo{volume}{87}(\bibinfo{number}{1}), pp. \bibinfo{pages}{93--116}.

\bibitemdeclare{article}{zhang2010survey}
\bibitem{zhang2010survey}
\bibinfo{author}{Ge-Xiang \surnamestart Zhang\surnameend} \&
  \bibinfo{author}{Lin-Qiang \surnamestart Pan\surnameend}
  (\bibinfo{year}{2010}): \emph{\bibinfo{title}{A survey of membrane computing
  as a new branch of natural computing}}.
\newblock {\slshape \bibinfo{journal}{Chinese journal of computers}}
  \bibinfo{volume}{33}(\bibinfo{number}{2}), pp. \bibinfo{pages}{208--214},
  \doi{10.3724/SP.J.1016.2010.00208}.

\bibitemdeclare{book}{zhang2017real}
\bibitem{zhang2017real}
\bibinfo{author}{Gexiang \surnamestart Zhang\surnameend},
  \bibinfo{author}{Mario~J \surnamestart P{\'e}rez-Jim{\'e}nez\surnameend} \&
  \bibinfo{author}{Marian \surnamestart Gheorghe\surnameend}
  (\bibinfo{year}{2017}): \emph{\bibinfo{title}{Real-life applications with
  membrane computing}}.
\newblock \bibinfo{volume}{25}, \bibinfo{publisher}{Springer},
  \doi{10.1007/978-3-319-55989-6}.

\end{thebibliography}
\end{document}